\documentclass[1p]{elsarticle}
\usepackage[OT4]{fontenc}
\usepackage{hyperref} 
\usepackage{graphicx} 
\usepackage{psfrag}
\usepackage{subfig}

\begin{document}
\begin{frontmatter}

\title{Paradox of integration - a computational model}

\author[agh]{Ma{\l}gorzata J. Krawczyk}  
\author[agh]{Krzysztof Ku{\l}akowski\corref{cor}}
\ead{kulakowski@fis.agh.edu.pl}

\address[agh]{AGH University of Science and Technology,
Faculty of Physics and Applied Computer Science,
al.~Mickiewicza~30, 30-059 Krak\'ow, Poland.
}

\cortext[cor]{Corresponding author}


\begin{abstract}
The paradoxical aspect of integration of a social group has been highlighted by Peter Blau ({\it Exchange and Power in Social Life}, Wiley and Sons, 1964). During the integration process, the group members simultaneously compete for social status and play the role of the audience. Here we show that when the competition prevails over the desire of approval, a sharp transition breaks all friendly relations. However, as was described by Blau, people with high status are inclined to bother more with acceptance of others; this is achieved by praising others and revealing her/his own weak points. In our model, this action smooths the transition and improves interpersonal relations.
\end{abstract}


\begin{keyword}
social systems \sep Heider dynamics \sep asymmetric relations \sep jammed states

\end{keyword}
\end{frontmatter}

\section{Introduction}

More than a half of century ago, Blau described a social phenomenon which he called ``paradox of integration'' \cite{blau64}. According to Blau, an integration of a social group includes two competing processes: attempts to appear attractive raise both attraction and repulsion.  While the former reaction is natural, the latter comes from the fear of being dominated. As a paradoxical consequence, most attractive persons can be rejected by the group. Having this in mind, persons both attractive and smart maintain their popularity by self-mockery and praising others. Up to our knowledge the effect remains unnoticed by social modellers, despite its importance as of a collective social phenomenon.\\ 

Taking the Blau description as granted, we intend to sharpen the picture of the paradox by developing its quantitative aspects. There is vast literature about dangers of quantitative social modeling, provided by both sociologists and  modellers themselves \cite{elias,helbing,edmonds}. Taking this into account, we are more attached to the internal logic of the social phenomenon, provided by the model, than to the calculated values of the model variables. We believe that the quantitative research should provide scenarios based on hypothetical ``what if'' assumptions. Below, attempts to attain higher status at expense of somebody else will be encoded symbolically as 'critique', and attempts to reach sympathy -- as 'praising'. We are going to consider two versions of the model, without and with the self-deprecating strategy. Although we do not expect this strategy to be absent in real societies, we hope that the counterfactual approach allows to identify its consequences.\\

According to both scenarios considered here, an agent $i$ praises or critiques another agent $j$, loosing or gaining her/his own status, respectively (for simplicity we write on males from now on). Simultaneously, the status of a praised agent $j$ increases, and the status of a criticized agent $j$ is reduced. These variations are balanced, however, by the acceptance of $i$ by the praised agent $j$, altogether with the acceptance of $i$ by those with the same status as $j$. On the contrary, a critique raises hostility towards $i$ of the criticized agent $j$ as well as of those with the same status as $j$. In the second scenario, agents adopt also the self-deprecating strategy. Then, their utility functions depend additionally on their actual status: if one's need of high status is already fulfilled, an agent is more prone to praise others \cite{blau64}. A reverse of this strategy is the shame-rage spiral \cite{scheff}; own status perceived as low is known to trigger aggression. \\

The goal of this work is to capture the collective character of the phenomenon. Coupling between agents is due to the variations of interpersonal relations, which involve all agents of the same status as the one who is praised or criticized. (A similar solidarity has been suggested by assuming that agents of the same size are prone to cooperate \cite{caram}.) This group reaction weakens, however, due to a creeping polarization with respect to status. Social polarization known to be ubiquitous \cite{pahl,turner}; its relation to our work is limited to the process of status formation in groups. Yet, even this limited aspect of the polarization is essential for the formation of group structure, group perception and identity formation \cite{ridgeway,cremer}. Although the polarization is not our founding assumption, it appears as a natural consequence of the modelled dynamics.  \\

In the next section, the algorithm is described in details. Sections 3 and 4 are devoted to our numerical results and their discussion, respectively. In the last section we note that the term ``paradox of integration'' has been used recently in a different meaning, and we discuss the mutual relation of these two phenomena. 

\section{Algorithm}

In a fully connected network of $N$ nodes, an agent is represented by a node. A social status $A_i$ is assigned to each agent $i$; initial values of those variables are small integers, 
selected randomly to be zero or $\pm 1$. Here, integer representation is chosen for its simplicity. The relations between agents are encoded in the form of an asymmetric matrix $x(i,j)=\pm 1$, with elements $+1$ (friendly) or $-1$ (hostile). The matrix element $x(i,j)$ specifies the relation of $i$ towards $j$.\\

In the first scenario, the simpler one out of the two considered here, the evolution proceeds as follows. At each time step, an ordered pair $(i,j)$ of different agents is selected randomly. (This means, that we select an agent $i$ and next we select an agent $j$, with the probabilities $1/N$ and $1/(N-1)$ respectively, the same for all agents.) The $i$-th agent evaluates  his utility function $f(i,j)$ if he praises or critiques the $j$-th agent. To do this, he needs to know the number $v(A_j)$ of agents with the same status as the $j$-th agent, including $j$ himself. The decision -- to praise or to critique -- is taken by checking the sign of $f(i,j)$ given by

\begin{equation}
 f(i,j)=-p+\frac{1-p}{N-1}v(A_j)
 \label{uti}
\end{equation}
Here, $p$ is the weight of the preference of status, and $1-p$ is the weight of the preference of acceptance. Once the former prevails, i.e. $f(i,j)<0$, the decision is to critique. Then, $x(k,i)$ is set to be $-1$ for all agents $k$ such that $A_k=A_j$. Next, the status $A_i$ is increased by one and the status  $A_j$ is reduced by one; note that the change of status concerns only two agents. In the opposite case, when $f(i,j)>0$, the decision is to praise. Then, all changes go quite the opposite: $x(k,i)$ is set to be $+1$ for all agents $k$ such that $A_k=A_j$, next $A_i \to A_i-1$ and $A_j \to A_j+1$. In this way, the mean value of the status is kept constant within the model; each increase of $A_j$ is accompanied by a decrease of $A_i$, and the opposite. This makes the status similar to IQ, with its mean value assumed to be 100 for each society. Actually, only relative variations of status are relevant.\\
\\

The second scenario is all the same; the only modification is that the agents' decisions include the dependence of the weight $p$ on the actual status of the decision-maker. As explained above, the mechanism is that $p$ decreases with $A_i$. Here we redefine the parameter $p$ as

\begin{equation}
p'_i=\frac{2p}{1+2^{A_i}}
 \label{pprim}
\end{equation}
what keeps $p'_i<1$ as long as $p<0.5$. This function captures both effects: the self-deprecating strategy and the shame-rage spiral. \\

\section{Results}

For both scenarios, we trace the mean value $<x>$ of all relations $x_{ij}$ against time $t$, averaged also over 100 realizations. This is an indicator of the kind of relations: friendly (positive) or hostile (negative). Also, keeping in mind that the mean value of the status remains constant, we are interested in its standard deviation. Initial values of the relations $x(i,j)$ do not influence the results, as they are forgotten in a few time steps. The model dynamics describes the process of differentation of status; then it makes sense to assume a narrow distribution of the initial values of $A_i$, as we do.\\

\begin{figure}[!hptb]
\begin{center}
\includegraphics[width=.89\columnwidth, angle=0]{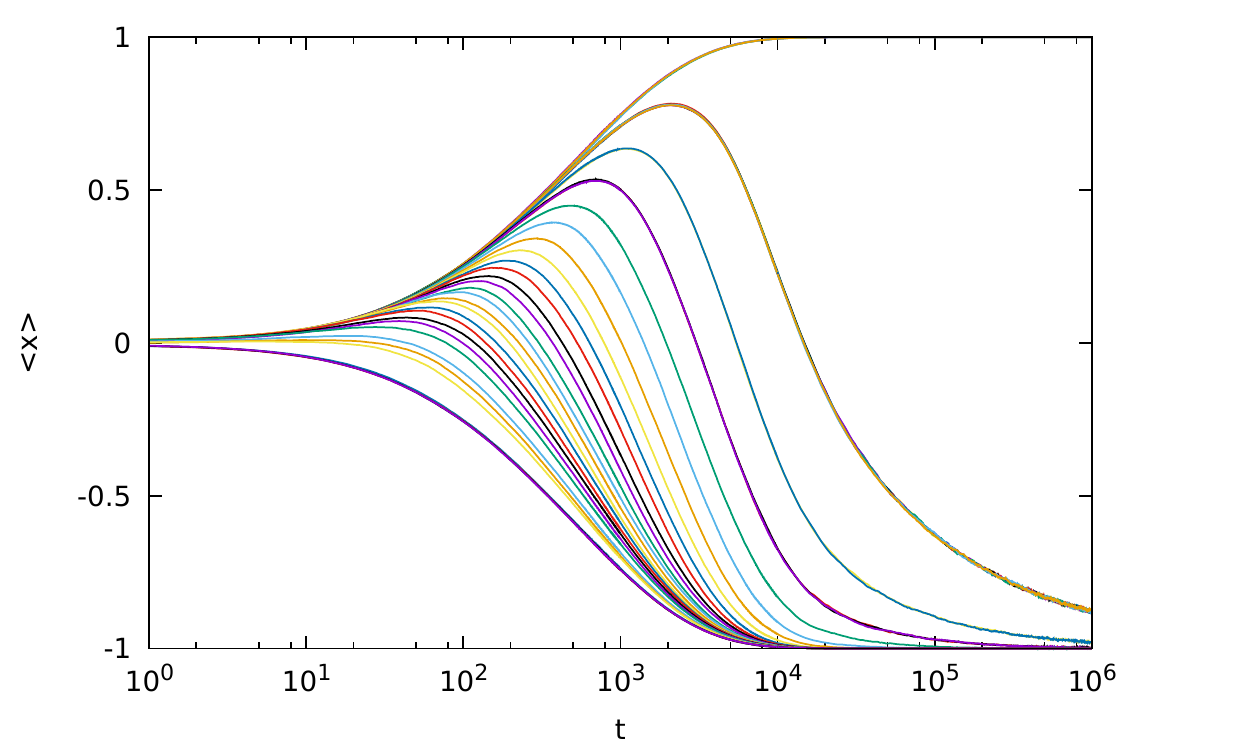}
\caption{The mean value of relations $<x>$ against time, for different values of the parameter $p$, calculated for the first scenario for $N=75$ and averaged over 100 simulations. The line tending to $+1$ includes the data for four values of $p$: 0.010, 0.011, 0.012 and 0.013. The remaining plots tend to -1, in the order from upper to lower curves when $p$ increases from 0.014 to 0.45.}
\label{fig1}
\end{center}
\end{figure}

\begin{figure}[!hptb]
\begin{center}
\includegraphics[width=.8\columnwidth, angle=0]{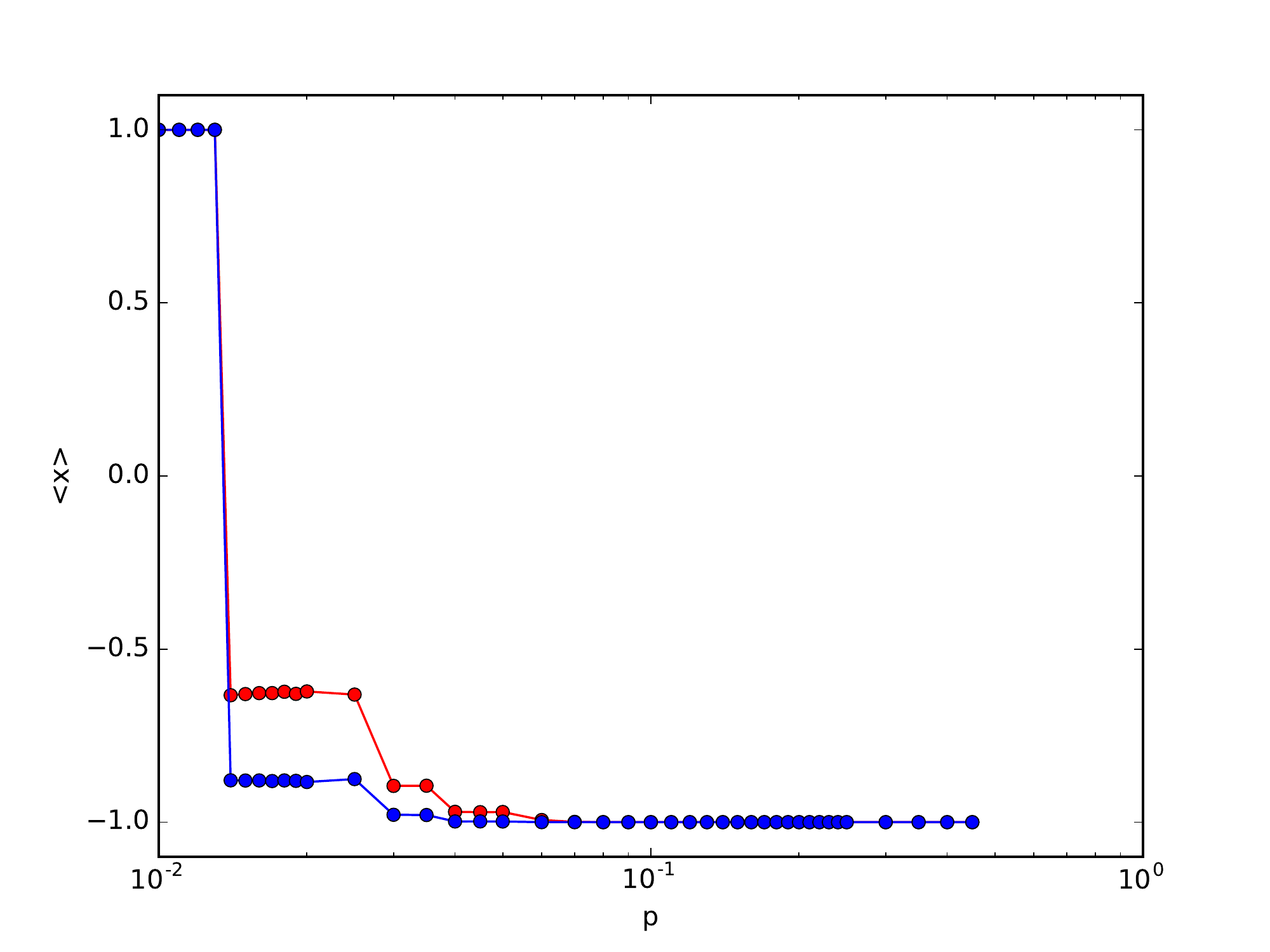}
\caption{The mean value of relations $<x>$ against the parameter $p$, after two different times of evolution ($t=10^5$ and $10^6$), calculated for the first scenario for $N=75$ and averaged over 100 simulations. The lower curve is for the longer time.}
\label{fig1a}
\end{center}
\end{figure}

\begin{figure}[!hptb] 
\begin{center}
\includegraphics[width=.8\columnwidth, angle=0]{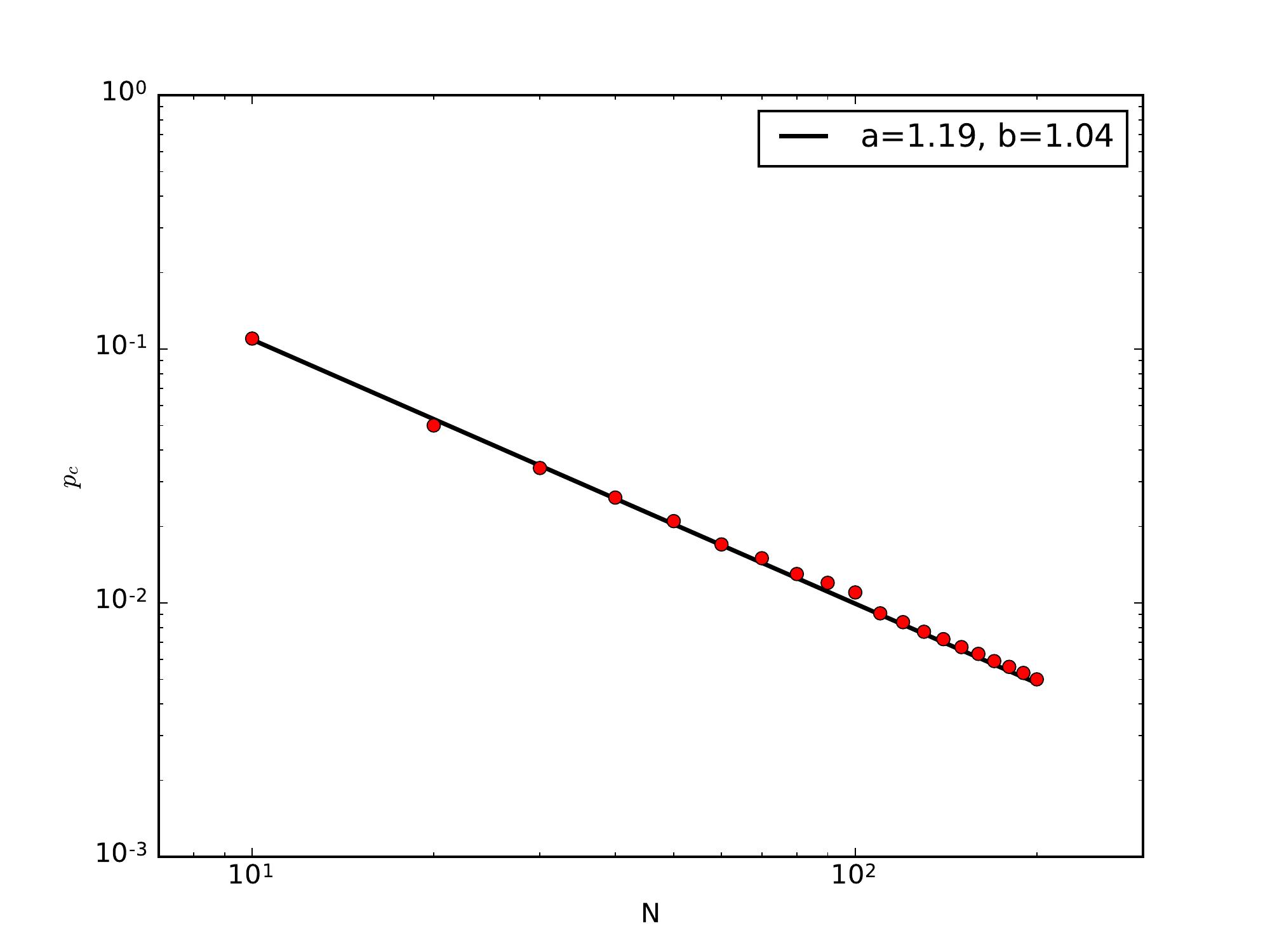}
\caption{The transition point $p_c$ as dependent on the system size, together with the fit to the function $aN^{-b}$. }
\label{fig1b}
\end{center}
\end{figure}

The results on $<x>$ obtained within the first scenario, where the weight $p$ is constant, are shown in Fig. (\ref{fig1}). It appears that the behaviour of $<x>$ sharply depends on the parameter $p$, and this dependence manifests only after some transient time. There is a critical value of $p$, denoted as $p_c$ from now on; for $p<p_c$, the mean relation tends to $+1$, while for $p>p_c$ it goes slowly to negative values. For $N=75$, we get $p_c=0.013$. This critical value is apparent, when we plot $<x>$ against $p$ after long time of calculations; such a plot is shown in Fig. (\ref{fig1a}). Further, the transition point $p_c$ is found to depend on the system size $N$; this dependence is shown in Fig. (\ref{fig1b}) together with the fit to the function $aN^{-b}$. The obtained parameters $a,b$ are both close to unity.\\

\begin{figure}[!hptb]
\begin{center}
\includegraphics[width=.89\columnwidth, angle=0]{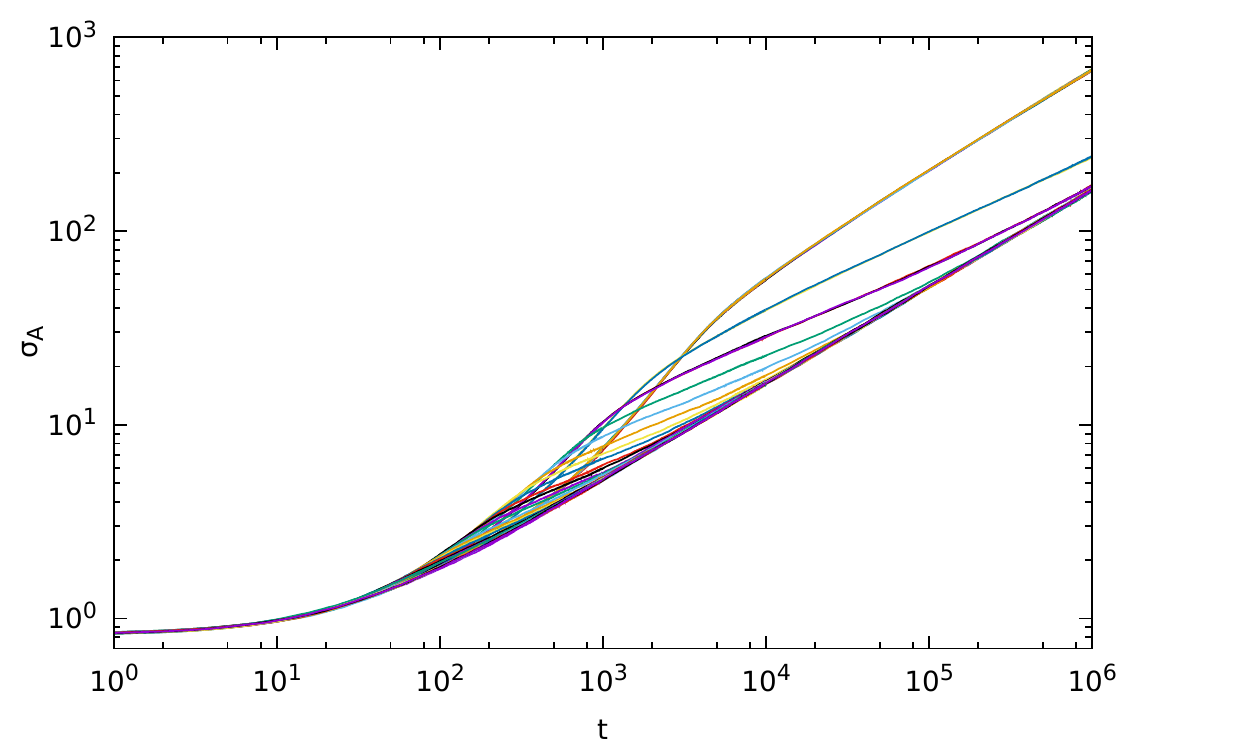}
\caption{The standard deviation $\sigma $ of status $A$ against time, for different values of the parameter $p$, calculated for the first scenario for $N=75$ and averaged over 100 simulations. For time $10^6$, the exponent $\alpha$ (defined by $\sigma _A^2 \propto t^{\alpha }) $ is close to one both for the maximal and minimal $\sigma _A$. The former is related to $p$ between 0.014 and 0.25. }
\label{fig2}
\end{center}
\end{figure}

The standard deviation of the status $\sigma_A$ is shown in Fig. (\ref{fig2}) as a function of time, for different values of $p$.  As a rule, the standard deviation increases without limits. The related exponent in the diffusion-like law $\sigma ^2 \propto t^\alpha$, measured after the transient time $t \approx 10^5$, is close to one. An exception is the range of $p$ between 0.03 and 0.05, where $\alpha$ is closer to 0.8; we deduce from Fig. (\ref{fig2}) that this behaviour is transient. In any case, as the distribution of $A$ gets wider, the numbers $v(A)$ decrease to one or zero; there is less and less of cases where there is more than one agent with a given status.\\

The results obtained within the second scenario are shown in Figs. ({\ref{fig3}) and ({\ref{fig4}), on $<x>$ and $\sigma_A$, respectively. Comparing the results with those in Figs. ({\ref{fig1}) and ({\ref{fig2}), we can identify the consequences of the application of the formula for $p'$ (Eq. \ref{pprim}). First, as shown if Fig. ({\ref{fig3}), the division of the plots $<x>(t)$ into two sets, tending asymptotically to $\pm 1$, is not visible; the dependence of $<x>$ on $p$ after long time of calculations ($10^6$ timesteps) remains continuous. On the other hand, although initially the plots for larger values of $p$ decrease similarly in Figs.  ({\ref{fig1}) and  ({\ref{fig3}), in the second scenario those decreasing functions turn back towards positive values. \\

The results shown in Fig. ({\ref{fig4}) indicate, that in the second scenario the standard deviation $\sigma _A$ remains much more limited. Here the difference between the plots for $p<p_c$ and those for $p>p_c$ is visible after some transient time about $10^4$ time steps: the former are of slightly but clearly higher values. Also, after the transient time $t\approx 10^5$ the exponent $\alpha $ is about 0.25 for $p<p_c$. Above $p_c$ it gradually tends to zero. We conclude that in the second scenario, the standard deviation of status is a better tool than $<x>$ to detect the transition.\\

\section{Discussion} 

After numerous variations of individual status of each agent it is unlikely that two agents have the same status. Hence after asymptotically long time, $v(A_j)$ is equal to one for all agents $j$. Then, the time dependence of $<x>$ in the first scenario can be interpreted in terms of a simple rule: if $-p+\frac{1-p}{N-1}$ is negative, $<x>$ decreases, and if $-p+\frac{1-p}{N-1}$ is positive, $<x>$ increases. Accordingly, the transition is at $p=p_c=1/N$. as observed numerically in Fig. {\ref{fig1b}}. On the other hand, in the second scenario there is no one common value of $p$, and that is why the transition is fuzzy there. However, as $<A>$ remains equal to zero and its distribution is symmetric and homogeneous ($v(A)=1$), the average value of $p'$ is\\

\begin{equation}
 <p'>=\frac{\sum_Av(A)p'(A)}{\sum_Av(A)}=p\frac{\sum_A\frac{2}{1+2^A}}{\sum_A1}=p
 \label{norm}
\end{equation}
and that is why $p_c=1/N$ in both scenarios. On the other hand, the drift of the status $A$ towards positive or negative values is damped by a reduction or an increase of $p'$, according to Eq. (\ref{pprim}). That is why the standard deviation of the status distribution is smaller in the second scenario.\\

Similar argument allows to interpret the stair-like behaviour of the function $<x>$ against $p$, shown in Fig. (\ref{fig1a}). During the time evolution of the relations, the contribution of pairs of agents with the same value of status gradually decreases. Yet this contribution is the same for the values of $p$ between $1/N$ and $2/N$. For $N=75$, $2/N=0.026(6)$; this is where the plateau of $<x>$ in Fig. (\ref{fig1a}) ends.\\

This part of discussion applies to the model regime after the transient time, when $v(A)$ is zero or one for most values of $A$. Then, the coupling between agents is reduced to interaction in pairs. As we see in Fig. (\ref{fig1}) it is only then when the decoupling appears. In this regime, the model is akin to the Bonabeau model \cite{bonabeau}; there, the standard deviation plays the role of the order parameter. At earlier times, most of the curves $<x>(t)$ increase even for $p>p_c$, because as long as $v>1$, $f(i,j)>0$ and the willingness to praise prevails. \\

The goal of this paper is to highlight the role of the self-deprecating strategy and of the accompanying shame-rage spiral. These roles are encoded in Eq. (\ref{pprim}). Without the coupling of the ambition to get high status (encoded in the parameter $p$) with the status itself (encoded as $A$), the transition from the phase 'all relations positive' to the phase 'most relations negative' is sharp and therefore hard to be predicted. \\

\begin{figure}[!hptb]
\begin{center}
\includegraphics[width=.89\columnwidth, angle=0]{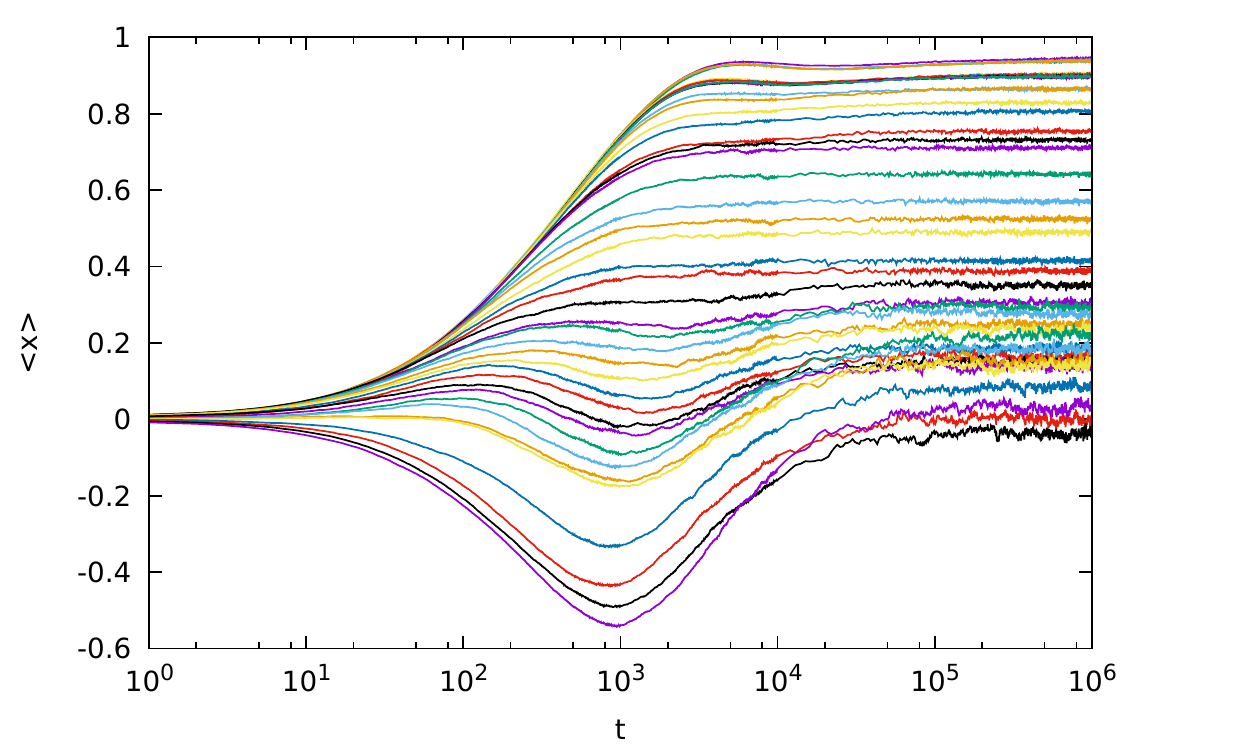} 
\caption{The mean value of relations $<x>$ against time, for different values of the parameter $p$, calculated for the second scenario for $N=75$ and averaged over 100 simulations. As $p$ increases from 0.010 to 0.45, $<x>$ decreases; the dependence is almost exactly monotonous.}
\label{fig3}
\end{center}
\end{figure}

\begin{figure}[!hptb]
\begin{center}
\includegraphics[width=.89\columnwidth, angle=0]{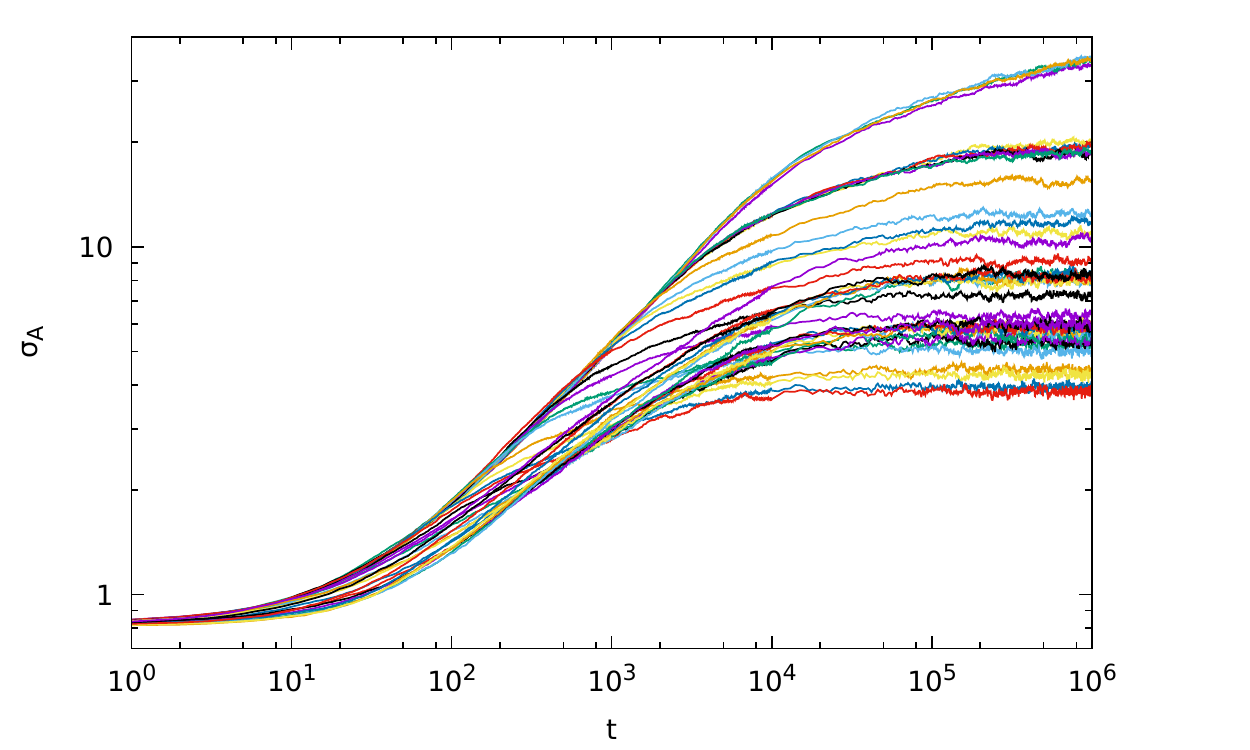}
\caption{The standard deviation $\sigma $ of status $A$ against time, for different values of the parameter $p$, calculated for the second scenario for $N=75$ and averaged over 100 simulations. At $t=10^6$, $\sigma _A$ is the largest for $p=$  0.010, 0.011, 0.012 and 0.013. Also, the exponent $\alpha$ (defined by $\sigma _A^2 \propto t^{\alpha }$) is close to 0.25 for these four plots. For higher $p$, $\alpha $ decreases monotonously to zero.  }
\label{fig4}
\end{center}
\end{figure}

We note that the assumption on the normalization of the term proportional to $1-p$ in Eq. (\ref{uti}) by $N-1$ is somewhat arbitrary. In fact, it is not clear how our decisions depend on the number of those who observe us. For instance, the authors of \cite{latane,nowak} argue that the influence of neighbours increases with their number less than linearly; the more neighbours, the smaller contribution of one neighbour. If we remove the above normalization, the transition remains sharp but it appears at $p_c=0.5$, independently on $N$. Between these two options, intermediate solutions are possible. Another point is to use the parameter $\gamma $ instead of 2 in Eq. (\ref{pprim}), which could read $p'=2p/(1+\gamma ^A)$. We expect that for $\gamma $ in the range (1,2) the results are intermediate between those obtained within the two scenarios.\\

\section{Final remarks}

The term 'paradox of integration' has been used recently \cite{teije} to describe the phenomenon that immigrants who are better educated and better integrated in the host society are more sensitive to ethnic discrimination. Hence, they can feel rejected more often than their less advanced compatriots. Here we intend to add a comment how these two different but identically termed phenomena interact.\\

It is useful to refer to the concept of hierarchy of needs by Maslow \cite{maslow}. We have to imagine two societies, the host native population (A) and the smaller society of immigrants (B). When these groups interact, the elite of B has a lower status than the elite of A. Hence their need of esteem is not fulfilled, and their use of the self-deprecating strategy makes no sense. As a consequence, elites of B are treated by their less integrated compatriots as conceited guys who show disrespect to their tradition and unsuccessfully crawl to A. This conflict can be more painful than the tension with A on its own.\\

\vspace{1cm}
\noindent
{\bf Acknowledgements}\\
  The work was partially supported by the PL-Grid Infrastructure.

\newpage

\vspace{1cm}
\noindent
{\bf References}\\


\end{document}